\documentclass[12pt,a4paper]{article}
 \usepackage{epsf}
\begin{document}
\title{The process $e^{+}e^{-}\rightarrow \bar{t}c$
in topcolor-assisted technicolor models}
\author{Chongxing Yue$^{(a,b)}$, Yuanben Dai$^{(b,c)}$,
 Qingjun Xu$^{b}$, Guoli Liu$^{b}$\\
{\small a: CCAST (World Laboratory) P.O. BOX 8730. B.J. 100080
P.R. China}\\
{\small b: College of Physics and Information
Engineering,}\\
 \small{Henan Normal University, Xinxiang  453002. P.R.China}
 \thanks{This work is supported by the National Natural Science
 Foundation of China(I9905004), the Excellent Youth Foundation of
  Henan Scientific Committee(9911), and Foundation of Henan Educational
  Committee.}
\thanks{E-mail:cxyue@public.xxptt.ha.cn}\\
{\small c: Institute of Theoretical Physics, Academia Sinica, P.O.
BOX 2735, B.J. 100080 P.R. China }}

\date{\today}
\maketitle
\begin{abstract}
\hspace{5mm}We consider the flavor changing neutral current(FCNC)
process $e^{+}e^{-}\rightarrow\bar{t}c$(or $t\bar{c}$) in the
context of topcolor-assisted technicolor(TC2) models. We find that
the contributions of charged scalars (technipions and top-pions)
and topcolor gauge bosons are negligibly small. The neutral
scalars (top-pion $\pi_{t}^{0}$ and top-Higgs $h_{t}^{0}$) can
give significant contributions to this process. With reasonable
values of the parameters in TC2 models, there will be several tens
of the observable $\bar{t}c$ events to be generated in the future
$e^{+}e^{-}$ linear colliders.
\end {abstract}

\vspace{1.0cm} \noindent {\bf PACS number(s)}: 14.65.Ha; 14.80.Cp

\newpage
\hspace{5mm}It is widely believed that the top quark, with a mass of the
 order of the electroweak scale, will be a sensitive probe into physics
 beyond the standard model(SM).The properties of the top quark could reveal
 information about flavor physics,
electroweak symmetry breaking(EWSB), as well as new physics beyond
the SM\cite{x1}. Much attention has been paid to the studies of
its production and decays. One of the most important tasks of
high-energy linear $e^{+}e^{-}$ colliders(LC) and hadron colliders
is the careful investigation of the production of top quark. As
the environment of the process $e^{+}e^{-}\rightarrow \bar{t}c$
(or $t\bar{c}$) is very clean, though the top quark has been found
at Tevatron, the LC still potentially offer a complement to those
of the hadron colliders studying the production of the top quark.

The search for flavor changing neutral current(FCNC) is one of the
most interesting possibilities to test the SM, with the potential
for either discovering or putting stringent bounds on new physics.
In the SM, there are no FCNC at tree-level and at one-loop level
they are GIM suppressed, which can offer a good place to test
quantum effects of the fundamental quantum field theory. In models
beyond SM, new particles may appear in the loop and have
significant contributions to FCNC processes. Therefore, FCNC
interactions give an ideal place to search for new physics. Any
positive observation of FCNC couplings deviated from that in SM
would unambiguously signal the presence of new physics. Thus,
searching for FCNC is also one of the major goals of the next
generation of high energy colliders, such as LC\cite{x2}.

The FCNC vertices $tcV$($V=\gamma,Z$) can be probed either directly in the
top-quark decays or indirectly in loops via top-charm associated production.
In this letter, we shall investigate the latter case in the FCNC process
 \begin{equation}
 e^{+}e^{-}\rightarrow\bar{t}c\hspace{3mm}(or\hspace{1mm}t\bar{c}).
 \end{equation}
Comparing $t$ quark rare decays where the momentum transfer
$q^{2}$ is limited, this process allows the large momentum
transfer, which is actually determined by the energies available
at LC. The process $e^{+}e^{-}\rightarrow\bar{t}c$ has some
advantages for probing higher dimension operators at large momenta
and striking kinetic signatures which are straight forward to
detect in the clean environment of LC. In models beyond SM which
can induce FCNC, there are large underlying mass scales and large
momentum transfer. So that these models are more naturally probed
via top-charm ($t\bar{c}$) associated production than top quark
rare decays\cite{x3}.

The production cross section of the FCNC process
$e^{+}e^{-}\rightarrow\bar{t}c$ in the SM has been calculated in
Ref.[4] and a lot of theoretical activity involving top-charm
production has been given within specific popular models beyond
the SM. For example, studies of $\bar{t}c$ production in multi
Higgs doublets models\cite{x5}, in supersymmetry with R-Parity
violation\cite{x6}, in models with extra vector-like
quarks\cite{x7}, in technicolor models\cite{x8}, and in a model
independent approach\cite{x3}. The aim of this letter is to point
out that the FC couplings predicted by topcolor-assisted
technicolor(TC2) models\cite{x9} can give significant
contributions to the FCNC process $e^{+}e^{-}\rightarrow\bar{t}c$,
which may be detected in the future LC experiments.

An important issue in high-energy physics is to understand the
mechanism of the mass generation. Given the large value of the top
quark mass and the sizable splitting between the masses of the top
and bottom quarks, it is natural to wonder whether $m_{t}$ has a
different origin from the masses of the other quarks and leptons
and there may be a common origin for EWSB and top quark mass
generation. Much theoretical work has been carried out in
connection to the top quark and EWSB. The TC2 models\cite{x9}, the
top see-saw models\cite{x10} and the flavor universal TC2
models\cite{x11} are three of such examples. Such type of models
generally predict a number of scalars with large Yukawa couplings
to the third generation. For example, TC2 models predict the
existence of scalars including the technipions ($\pi^{0}$,
$\pi^{\pm}$; $\pi^{0,a}$, $\pi^{\pm,a}$) in the technicolor
sector, and the top-pions ($\pi_{t}^{0}$, $\pi_{t}^{\pm}$) and the
top-Higgs boson $h_{t}^{0}$ in the topcolor sector. Ref.[8] has
considered the contributions of the scalars predicted by the
technicolor model with a massless scalar doublet\cite{x12}. The
results show that the cross section of the FCNC process
$e^{+}e^{-}\rightarrow\bar{t}c$ is of the order of
$10^{-3}-10^{-5}fb$, which is too small to be detected in the
future LC experiments. Further, Ref.[8] only considered the
contributions of charged scalars to this process via the coupling
$s^{\pm}u_{i}d_{j}$.

It is an important feature of TC2 models that the neutral scalars
can induce the tree-level FC scalar couplings. It has been shown
that the FC scalar couplings can give distinct new flavor-mixing
phenomena which may be tested at both low- and high-energy
experiments\cite{x13,x14}. Thus, in this letter, we will mainly
consider the contributions of the FC scalar coupling $s\bar{t}c$
($s$ is $\pi_{t}^{0}$ or $h_{t}^{0}$) to the FCNC process
$e^{+}e^{-}\rightarrow\bar{t}c$ in the context of TC2 models and
see whether the signals of TC2 models can be tested via the
process $e^{+}e^{-}\rightarrow\bar{t}c$ in the future LC
experiments. Our results show that the contributions of charged
scalars and the topcolor gauge bosons to the FCNC process
$e^{+}e^{-}\rightarrow\bar{t}c$ are all negligibly small. While
the neutral scalars can give significant contributions which may
be detected in the future LC experiments.

For TC2 models, the underlying interactions, topcolor
interactions, are nonuniversal and therefore do not possess a GIM
mechanism. This is an essential feature of this kind of models due
to the need to single out the top quark for condensate. This
nonuniversal gauge interactions result in FCNC vertices when one
writes the interactions in the quark mass eigenbasis. Thus, the
top-pions have large Yukawa coupling to the third family quarks
and can induce the new FC scalar couplings including the $t-c$
transitions for the neutral top-pion $\pi_{t}^{0}$\cite{x9,x13}:
\begin{eqnarray}
\nonumber
\frac{m_{t}}{\sqrt{2}F_{t}}\frac{\sqrt{\nu_{w}^{2}-F_{t}^{2}}}
{\nu_{w}}&[&iK_{UR}^{tt}K_{UL}^{tt*}\bar{t_{L}}t_{R}\pi_{t}^{0}+
\sqrt{2}K_{UR}^{tt*}K_{DL}^{bb}\bar{t_{R}}b_{l}\pi_{t}^{+}+\\
&&iK_{UR}^{tc}K_{UL}^{tt*}\bar{t_{L}}c_{R}\pi_{t}^{0}+
\sqrt{2}K_{UR}^{tc*}K_{DL}^{bb}\bar{c_{R}}b_{L}\pi_{t}^{+}+h.c.],
\end{eqnarray}
where $F_{t}$ is the top-pion decay constant and
$\nu_{w}=\nu/\sqrt{2}=174GeV$. It has been shown that the values
of coupling parameters can be taken as:
\begin{equation}
K_{UL}^{tt}=K_{DL}^{bb}=1, \hspace{5mm}  K_{UR}^{tt}=1-\epsilon,
\hspace{5mm} K_{UR}^{tc}\leq \sqrt{2\epsilon-\epsilon^{2}},
\end{equation}
with a model-dependent parameter $\epsilon\ll 1$. In the following
calculation, we will take
$K_{UR}^{tc}=\sqrt{2\epsilon-\epsilon^{2}}$ and take $\epsilon$ as
a free parameter.

The relevant Feynman diagrams for the contributions of the neutral
top-pion $\pi_{t}^{0}$ to the process
$e^{+}e^{-}\rightarrow\bar{t}c$ via the FC coupling are shown in
Fig.1. Using Eq.(2) and other relevant Feynman rules, we can give
the effective vertices $Z\bar{t}c$ and $\gamma\bar{t}c$ arising
from $\pi_{t}^{0}$:
\begin{equation}
\Lambda_{Z\bar{t}c}^{\mu}=ie\left[\gamma^{\mu}(F_{1Z}+F_{2Z}\gamma^{5})+
p_{t}^{\mu}(F_{3Z}+F_{4Z}\gamma^{5})+p_{c}^{\mu}(F_{5Z}+F_{6Z}\gamma^{5})
\right],
\end{equation}
\begin{equation}
\Lambda_{\gamma\bar{t}c}^{\mu}=\Lambda_{Z\bar{t}c}^{\mu}|_{F_{iZ}\rightarrow
F_{i\gamma}},\hspace{5mm}F_{i\gamma}=F_{iZ}|_{v_{t}=\frac{2}{3},
\hspace{2mm}a_{t}=0}.
\end{equation}
The form factors $F_{iV}$ are expressed in terms of two- and
three-point scalar integrals\cite{x16}. The expressions of
$F_{iV}$ are given in Appendix.

The total cross section can be written as:
\begin{equation}
\sigma=\frac{s-m_{t}^{2}}{32\pi s^{2}}\int_{0}^{\pi}\overline{|M|^{2}}
\sin\theta d\theta,
\end{equation}
with
\begin{equation}
\overline{|M|^{2}}=\frac{1}{4}|M|_{Z}^{2}+\frac{1}{4}|M|_{\gamma}^{2}
+\frac{1}{2}ReM_{Z}M_{\gamma}^{+}.
\end{equation}

The production cross section $\sigma$ contributed by the
$\pi_{t}^{0}$ is plotted in Fig.2 as a function of the top-pion
mass $m_{\pi_{t}}$ for the center-of-mass energy $\sqrt{s}=500GeV$
and three of the parameter $\epsilon$. From Fig.2 we can see that
the FC scalar coupling $\pi_{t}^{0}\bar{t}c$ could indeed give
significant contributions to the process
$e^{+}e^{-}\rightarrow\bar{t}c$. The value of the production cross
section $\sigma$ increase with $\epsilon$ and $m_{\pi_{t}}$
increasing. For $m_{\pi_{t}}=300GeV$, the cross section $\sigma$
varies between $0.05fb$ and $0.12fb$ for the parameter $\epsilon$
in the range of $0.03-0.08$. For $\epsilon=0.05$, $\sigma$ varies
between $0.076fb$ and $0.105fb$ for the top-pion mass
$m_{\pi_{t}}$ in the range of $200GeV-450GeV$.

To see the effects of the center-of-mass energy on the production
cross sections, we plot the $\sigma$ versus $\sqrt{s}$ in Fig.3
for $m_{\pi_{t}}=300GeV$ and three values of the parameter
$\epsilon$. One can see from Fig.3 that $\sigma$ increases with
$\sqrt{s}$ increasing for $\sqrt{s}\leq480GeV$ and decreases with
$\sqrt{s}$ increasing for $\sqrt{s}\geq480GeV$. For
$\sqrt{s}=480GeV$, the production cross section $\sigma$ reaches
the maximum value $\sigma_{max}=0.122fb$.

TC2 models also predict the existence of the neutral CP-even
state, called top-Higgs boson $h_{t}^{0}$, which is a $t\bar{t}$
bound and analogous to the $\sigma$ particle in low energy QCD.
Its mass can be estimated in the Nambu- Jona-Lasinio (NJL) model
in the large $N_{C}$ approximation and is found to be of the order
of $m_{h_{t}}\approx2m_{t}$\cite{x13}. The main difference between
$\pi_{t}^{0}$ and $h_{t}^{0}$ is that $h_{t}^{0}$ can couple to
gauge boson pairs $W W$ and $Z Z$ at the tree-level, which is
similar to that of the SM Higgs boson $H^{0}$. However, the
coupling coefficients of the couplings $h_{t}^{0}WW$ and
$h_{t}^{0}ZZ$ are suppressed by the factor $\frac{F_{t}}{\nu_{w}}$
with respect to that of $H^{0}$. Thus, the contributions of the
top-Higgs boson $h_{t}^{0}$ to the FCNC process
$e^{+}e^{-}\rightarrow\bar{t}c$ are similar to that of
$\pi_{t}^{0}$. Our numerical results are shown in Fig.4 for
$m_{h_{t}}=300GeV$. From Fig.4 we can see that $\sigma$ varies
between $0.042fb$ and $0.184fb$ for $\sqrt{s}$ in the range of
$300GeV-1000GeV$, which is larger than the production cross
section generated by $\pi_{t}^{0}$. Combining the contributions of
$h_{t}^{0}$ to the cross section $\sigma$ with that of the
top-pion $\pi_{t}^{0}$, the total cross section $\sigma_{total}$
given by the neutral scalars is in the range of $0.014fb- 0.35fb$
in most of the parameter space in TC2 models. For $\epsilon=0.08$,
$\sqrt{s}=480GeV$ and $m_{\pi_{t}}=m_{h_{t}}=450GeV$, the maximum
value of $\sigma_{total}$ can reach $0.35fb$.

The possible observable final states for the FCNC process
$e^{+}e^{-}\rightarrow\bar{t}c$ are $\bar{b}cj_{1}j_{2}$, where
$j_{1}$ and $j_{2}$ are light jets coming from $t\rightarrow
bW^{+}$ followed by $W^{+}\rightarrow u\bar{d}$ or $c \bar{s}$,
and $ b\bar{c}l^{+}\nu_{l}$, which comes from $t\rightarrow
bW^{+}\rightarrow bl^{+}\nu_{l}$ with $l=e$, $\mu$ or $\tau$.
These two final states occur with branching of $2/3$ and $1/3$,
respectively. The leading SM background of the FCNC process
$e^{+}e^{-}\rightarrow\bar{t}c$ mainly comes from $W$ pair
production as well as from $W$ bremsstrahlung in
$e^{+}e^{-}\rightarrow W+$ 2-jets\cite{x3}. We define the
background-free observable cross section $\bar{\sigma_{\bar{t}c}}$
as the effective cross section including $b$-tagging efficiency
($\epsilon_{b}$) and top quark reconstruction efficiency
($\epsilon_{t}$):
$\bar{\sigma_{\bar{t}c}}=\epsilon_{b}\epsilon_{t}\sigma_{\bar{t}c}$.

In order to estimate the number of the observable events, we
assume $\epsilon_{b}=60\%$ and $\epsilon_{t}=80\%$ as done by
Bar-Shalom and J. Wudka in Ref[3]. We consider two future
$e^{+}e^{-}$ collider scenarios: a LC with $\sqrt{s}=500GeV$ and a
yearly integrated luminosity of $L=100fb^{-1}$ and a LC with
$\sqrt{s}=1000GeV$ and $L=200fb^{-1}$. The yearly observable
production events can be easily calculated. Our numerical results
are shown in Fig.5. In Fig.5 we have assumed $m_{\pi_{t}}\approx
m_{h_{t}}= m $ and take $\epsilon=0.08$. We can see from Fig.5
that there will be several tens of the observable $ \bar{t}c $
events to be generated in the future LC experiments. The number of
the observable $ \bar{t}c $ events generated in a LC with
$\sqrt{s}=500GeV$ is larger than that generated in a LC with
$\sqrt{s}=1000GeV$. Furthermore, for $ 350GeV< m < 500GeV $, the
number generated in a LC with $\sqrt{s}=500GeV$ increases with $ m
$ increasing. Thus, it is likely that the signals of TC2 models
can be more easily detected at a LC with $\sqrt{s}=500GeV$ than at
a LC with $\sqrt{s}=1000GeV$.

If the masses of the neutral scalars are smaller than $350GeV$, $
\bar{t}c $ and $gg$ are the dominant decay modes of the scalars
$\pi_{t}^{0}$ and $h_{t}^{0}$. In this case, these new particles
have significant contributions to the top-charm associated
production at hadron colliders. This has been extensively studied
by Burdman\cite{x13}. His results show that the production cross
section of the process $gg\rightarrow\pi_{t}^{0}
(h_{t}^{0})\rightarrow\bar{t}c$ is very large. There will be
several thousands of $\bar{t}c$ events to be generated at the LHC,
which are larger than the number of $\bar{t}c$ events produced at
the future LC experiments. Thus, it is possible that the signals
of TC2 models can be more easily detected at the LHC than at the
LC experiments. However, the cross section of the process
$gg\rightarrow\pi_{t}^{0}(h_{t}^{0})\rightarrow\bar{t}c$ decrease
by two or even more orders for $m_{h_{t}(\pi_{t})}>350GeV$.
Furthermore, we must separate the signals from the large
background before observation of the neutral scalars at the LHC.
Thus, it is possible that the neutral scalars are more easily
detected at the LC than at the LHC for
$m_{h_{t}(\pi_{t})}>400GeV$.

To solve the phenomenological difficulties of the traditional TC
models, TC2 models\cite{x9} were proposed by combining technicolor
interactions with the topcolor interactions for the third
generation at the scale about $1TeV$. Thus, TC2 models predict
number of technipions in the technicolor sector. These new
particles also have contributions to the FCNC process
$e^{+}e^{-}\rightarrow\bar{t}c$ via the flavor diagonal couplings
$s^{\pm}u_{i}d_{j}$. The graphs of $e^{+}e^{-}\rightarrow\bar{t}c$
through charged technipions and top-pions loop are shown in Fig.6.
Our calculation results show that the contributions of charged
scalars to the FCNC process $e^{+}e^{-}\rightarrow\bar{t}c$ are
smaller than that of the neutral scalars. In the most of parameter
space, the production cross section is only in the range of
$10^{-2}-10^{-5}fb$, which can not be observed in the future LC
experiments.

In TC2 models, to maintain electroweak symmetry between top and
bottom quarks and yet to not generate $m_{b}\approx m_{t}$, the
topcolor gauge group is usually taken to be a strongly coupled
$SU(3)\times U(1)$. The $U(1)$ provides the difference that cause
only top quark to condensate. Thus, TC2 models predict the
existence of topcolor gauge bosons $B_{\mu}^{A}$ and an extra
$U(1)_{Y}$ gauge boson $Z^{\prime}$. Tree-level FCNC for the
topcolor gauge bosons $B_{\mu}^{A}$ and $Z^{\prime}$ are generated
when quarks fields are rotated to the mass eigenstate basis. The
couplings of $Z^{\prime}$ are nonuniversal and stronger for the
third generation, yielding potentially large top-charm couplings.
Ref.[15] has shown that the FC coupling can indeed give
significant contributions to the FCNC process
$e^{+}e^{-}\rightarrow\bar{t}c$. For
$M_{Z^{\prime}}=M_{B^{\prime}}=1TeV$, the production cross section
is of the order of $10^{0}-10^{-1}fb$. However, Ref.[17] has
pointed out $B\bar{B}$ mixing provides strong lower bounds on the
masses of $B_{\mu}^{A}$ and $Z^{\prime}$ bosons, i.e. there must
be larger than 4 $TeV$. In this case, the cross section $\sigma$
of the process $e^{+}e^{-}\rightarrow\bar{t}c$ given by topcolor
gauge bosons is smaller than $10^{-2}fb$. Recently, E. H.
Simmons\cite{x18} has shown that the B-meson mixing can also given
lower bound on the mass of the $Z^{\prime}$ boson predicted by the
flavor-universal TC2 models\cite{x11} or non-commuting extended
technicolor(NCETC) models\cite{x19}. However, the bound is much
weak, i.e. $M_{Z^{\prime}}>373GeV$. If we assume the mass of
$Z^{\prime}$ predicted by the flavor-universal TC2 or NCETC models
approximately equals to $500GeV$, then the cross section $\sigma$
given by $Z^{\prime}$ is of the order of $10^{0}-10^{-1}fb$ in
most of the parameter space. Thus, the $Z^{\prime}$ gauge boson
predicted by these models can give significant contributions to
the FCNC process $e^{+}e^{-}\rightarrow\bar{t}c$, which may be
detected in the future LC experiments.

To completely avoid the problems, such as triviality and
unnaturalness arising from the elementary Higgs field, various
kinds of dynamical EWSB mechanisms have been proposed and among
which TC2 theory is an attractive idea. TC2 models predict number
of new scalars and gauge bosons. In this letter, we calculated the
contributions of these new particles to the FCNC process
$e^{+}e^{-}\rightarrow\bar{t}c$ in the framework of TC2 models.
Our results shown that the contributions of charged scalars and
topcolor gauge bosons to this process is very small. However, the
neutral top-pion $\pi_{t}^{0}$ and top-Higgs $h_{t}^{0}$ can give
significant contributions to the FCNC process
$e^{+}e^{-}\rightarrow\bar{t}c$. With reasonable value of the
parameters, the production cross section $\sigma$ can reach
$0.35fb$. There will be several tens of the observable $\bar{t}c$
events to be generated in the LC experiments with $\sqrt{s}=
500GeV$. Thus, the possible signals of $\pi_{t}^{0}$ and
$h_{t}^{0}$ may be observed in the future LC experiments via this
process.

\newpage
{\bf Appendix: The form factors in the effective vertices $Z\bar{t}c$ and
$\gamma\bar{t}c$ for the neutral
top-pion $\pi_{t}^{0}$:}\\
\hspace*{6mm}$F_{1Z}=g[B_{0}+m_{\pi_{t}}^{2}C_{0}-2C_{24}+m_{t}^{2}(C_{11}-C_{12})-B_{0}^{*}-B_{1}^{\prime}]v_{t}$,\\
\hspace*{6mm}$F_{2Z}=g[B_{0}+m_{\pi_{t}}^{2}C_{0}-2C_{24}-m_{t}^{2}(C_{11}-C_{12})+B_{0}^{*}+B_{1}^{\prime}-4\overline{C_{24}}]a_{t}$,\\
\hspace*{6mm}$F_{3Z}=2m_{t}g(C_{21}+C_{22}-2C_{23})v_{t}$,\\
\hspace*{6mm}$F_{4Z}=2m_{t}g(-C_{21}-C_{22}+2C_{23}-2\overline{C_{22}}+\overline{C_{12}}+\overline{C_{0}})a_{t}$,\\
\hspace*{6mm}$F_{5Z}=2m_{t}g(C_{22}-C_{23}+C_{12})v_{t}$,\\
\hspace*{6mm}$F_{6Z}=2m_{t}g(-C_{22}+C_{23}+C_{12}-2\overline{C_{22}}+3\overline{C_{12}}+2\overline{C_{23}}-2\overline{C_{11}}-\overline{C_{0}})a_{t}$.\\
Here $g=\frac{1}{16\pi^{2}}[\frac{m_{t}}{\sqrt{2}F_{t}}\frac{\sqrt{\nu_{w}^{2}-F_{t}^{2}}}{\nu_{w}}]^{2}K_{UR}^{tc}K_{UL}^{tt*}$,
$v_{t}=\frac{1}{4S_{W}C_{W}}(1-\frac{8}{3}S_{W}^{2})$, $a_{t}=\frac{1}{4S_{W}C_{W}}$.
The expressions of two- and three-point scalar integrals $B_{n}$ and $C_{ij}$
in this paper are\cite{x16}:\\
\hspace*{6mm}$B_{n}=B_{n}(-\sqrt{s}, m_{t}, m_{t})$,\hspace{5mm}$B^{*}_{n}=B_{n}(-p_{c}, m_{\pi_{t}}, m_{t})$,\\
\hspace*{6mm}$B^{\prime}_{n}=B_{n}(-p_{t}, m_{\pi_{t}}, m_{t})$,\hspace{18mm} $C_{ij}=C_{ij}(p_{t}, -\sqrt{s}, m_{\pi_{t}}, m_{t}, m_{t})$,\\
\hspace*{6mm}$C_{0}=C_{0}(p_{t}, -\sqrt{s}, m_{\pi_{t}}, m_{t}, m_{t})$,\hspace{5mm} $\overline{C}_{ij}=C_{ij}(-p_{c}, \sqrt{s}, m_{t}, m_{\pi_{t}}, m_{\pi_{t}})$,\\
\hspace*{6mm}$\overline{C}_{0}=C_{0}(-p_{c}, \sqrt{s}, m_{t}, m_{\pi_{t}}, m_{\pi_{t}})$.

\newpage
\vskip 2.0cm
\begin{center}
{\bf Figure captions}
\end{center}
\begin{description}
\item[Fig.1:]Feynman diagrams for the contributions of the neutral
top-pion $\pi_{t}^{0}$ to the process $e^{+}e^{-}\rightarrow\bar{t}c$.
The dashed lines are the top-pion inner lines.
\item[Fig.2:]The production cross section $\sigma$ contributed by
$\pi_{t}^{0}$ as a function of the mass $m_{\pi_{t}}$ for $\sqrt{s}=500GeV$
and $\epsilon=0.03$(solid line), $\epsilon=0.05$(short dashed line),
$\epsilon=0.08$(dot-dashed line).
\item[Fig.3:]The production cross section $\sigma$ contributed by
$\pi_{t}^{0}$ as a function of $\sqrt{s}$ for $m_{\pi_{t}}=300GeV$
and $\epsilon=0.03$(solid line), $\epsilon=0.05$(short dashed line),
$\epsilon=0.08$(dot-dashed line).
\item[Fig.4:]The production cross section $\sigma$ contributed by
$h_{t}^{0}$ as a function of $\sqrt{s}$ for $m_{h_{t}}=300GeV$
and $\epsilon=0.03$(solid line), $\epsilon=0.05$(short dashed line),
$\epsilon=0.08$(dot-dashed line).
\item[Fig.5:]The yearly observable total events contributed by
$h_{t}^{0}$ and $\pi_{t}^{0}$ as a function of $m$ for $\epsilon=0.08$
and $\sqrt{s}=500GeV$(solid line), $\sqrt{s}=1000GeV$(short-dashed line).
Here we assume $m_{h_{t}}=m_{\pi_{t}}=m$.
\item[Fig.6:]Feynman diagrams for the contributions of the charged
technipions and top-pions to the process $e^{+}e^{-}\rightarrow\bar{t}c$.
The dashed lines are the charged technipions or top-pions inner lines.
\end{description}

\newpage
\begin{figure}[pt]
\begin{center}
\begin{picture}(300,250)(0,0)
\put(-50,0){\epsfxsize160mm\epsfbox{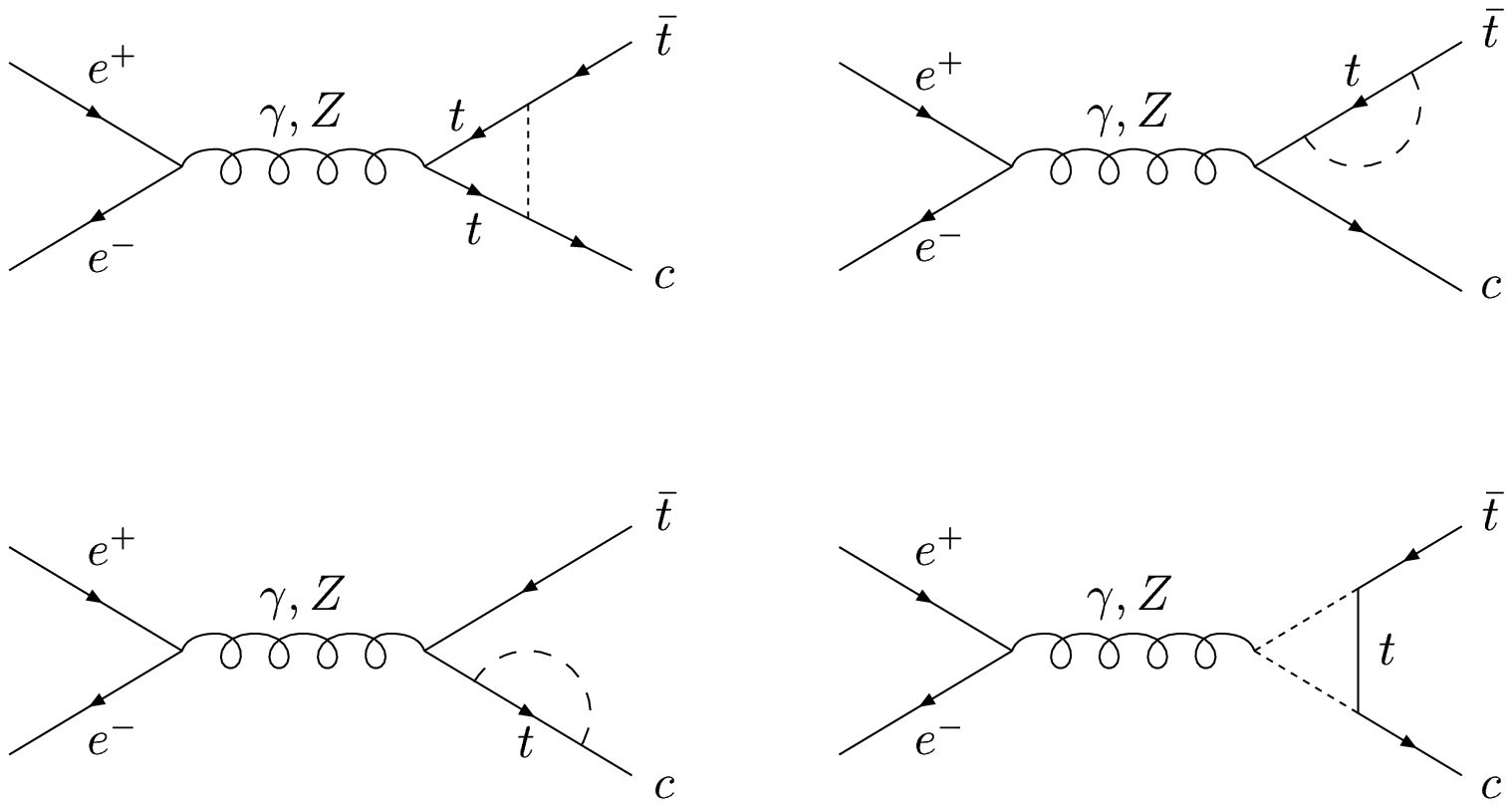}}\put(150,100){Fig.1}
\end{picture}
\end{center}
\end{figure}
\begin{figure}[hb]
\begin{center}
\begin{picture}(250,200)(0,0)
\put(-50,0){\epsfxsize120mm\epsfbox{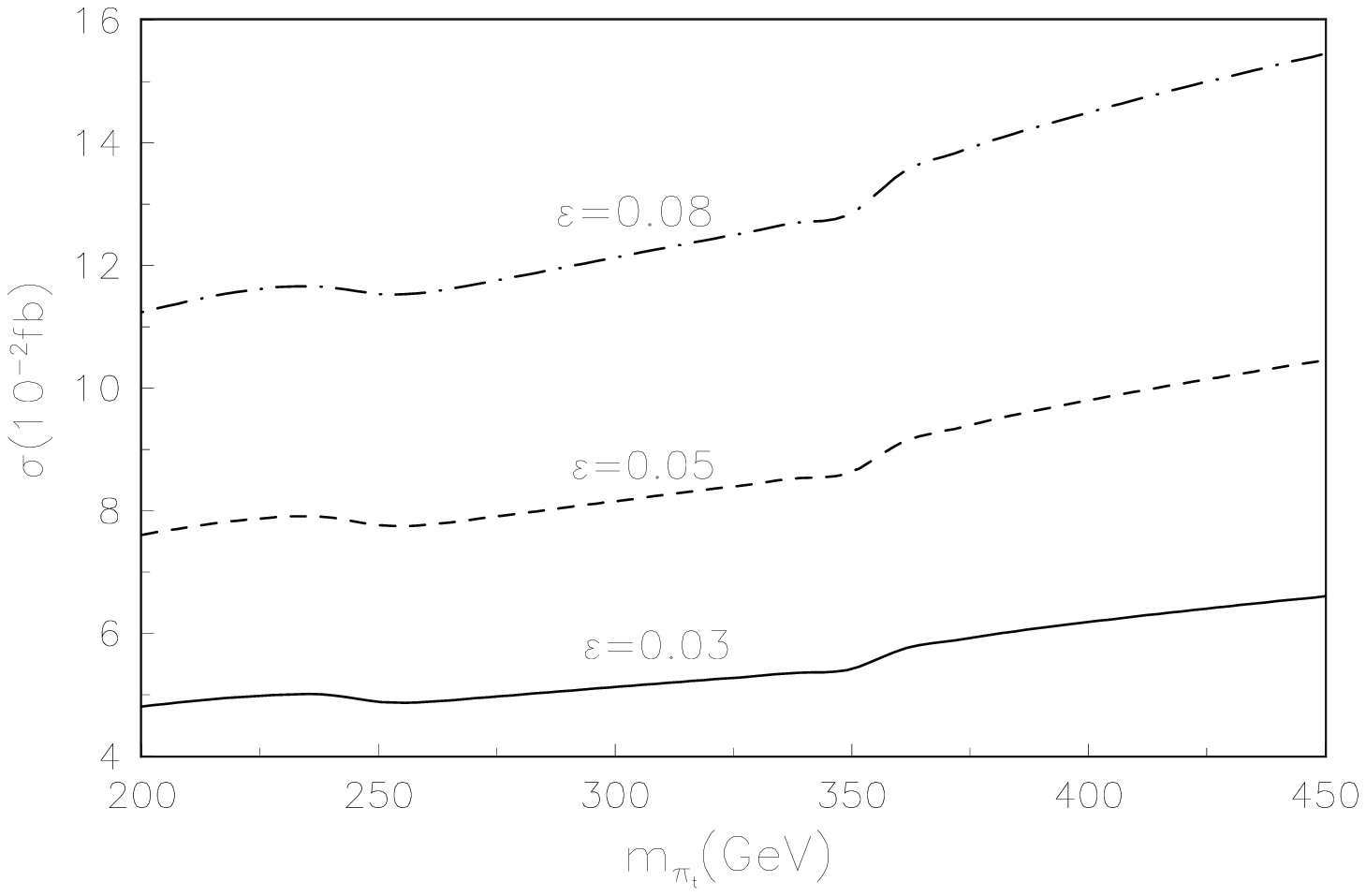}}\put(120,-10){Fig.2}
\end{picture}
\end{center}
\end{figure}
\newpage
\begin{figure}[pt]
\begin{center}
\begin{picture}(250,200)(0,0)
\put(-50,0){\epsfxsize120mm\epsfbox{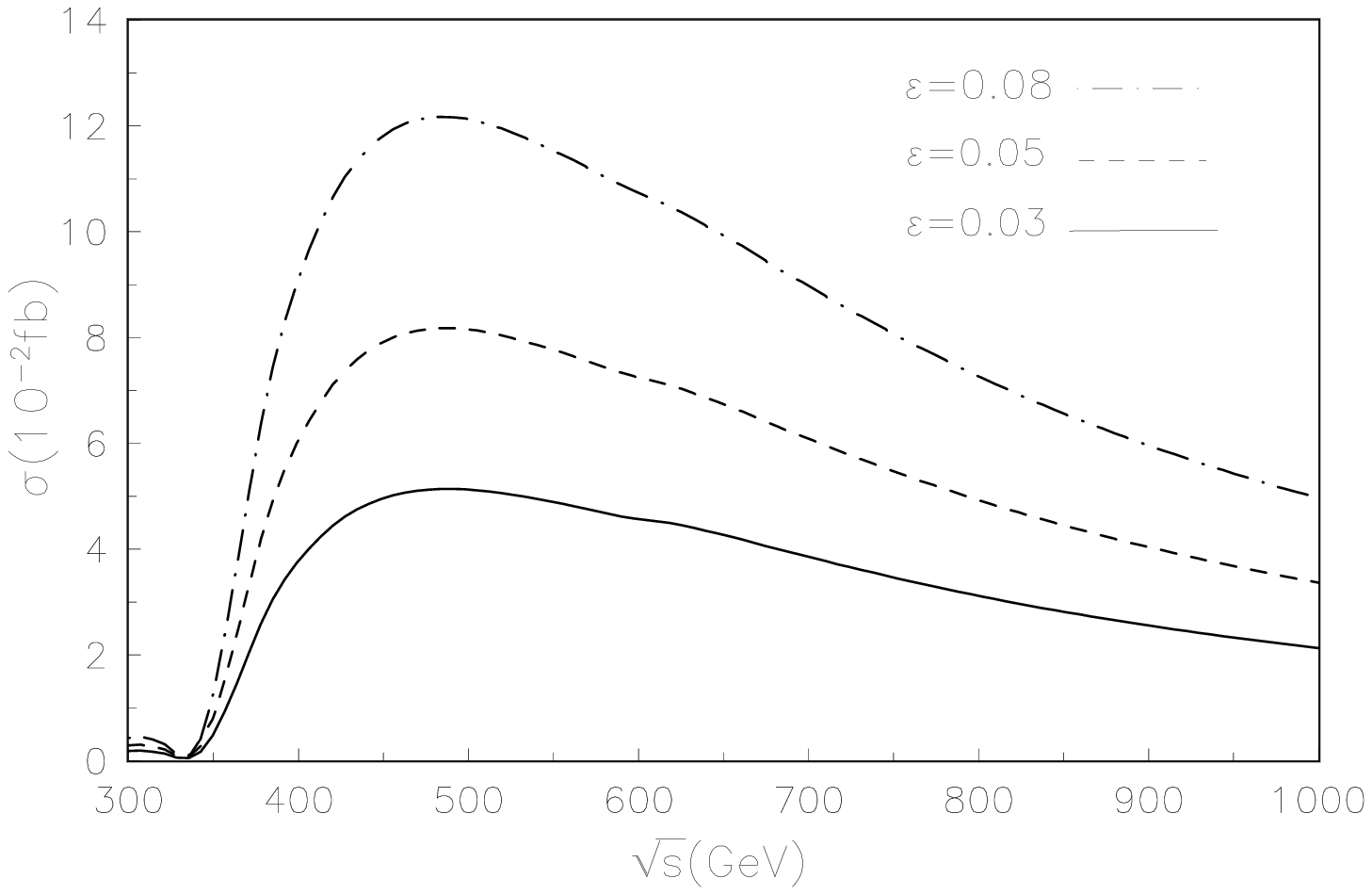}}\put(120,-10){Fig.3}
\end{picture}
\end{center}
\end{figure}
\begin{figure}[hb]
\begin{center}
\begin{picture}(250,200)(0,0)
\put(-50,0){\epsfxsize120mm\epsfbox{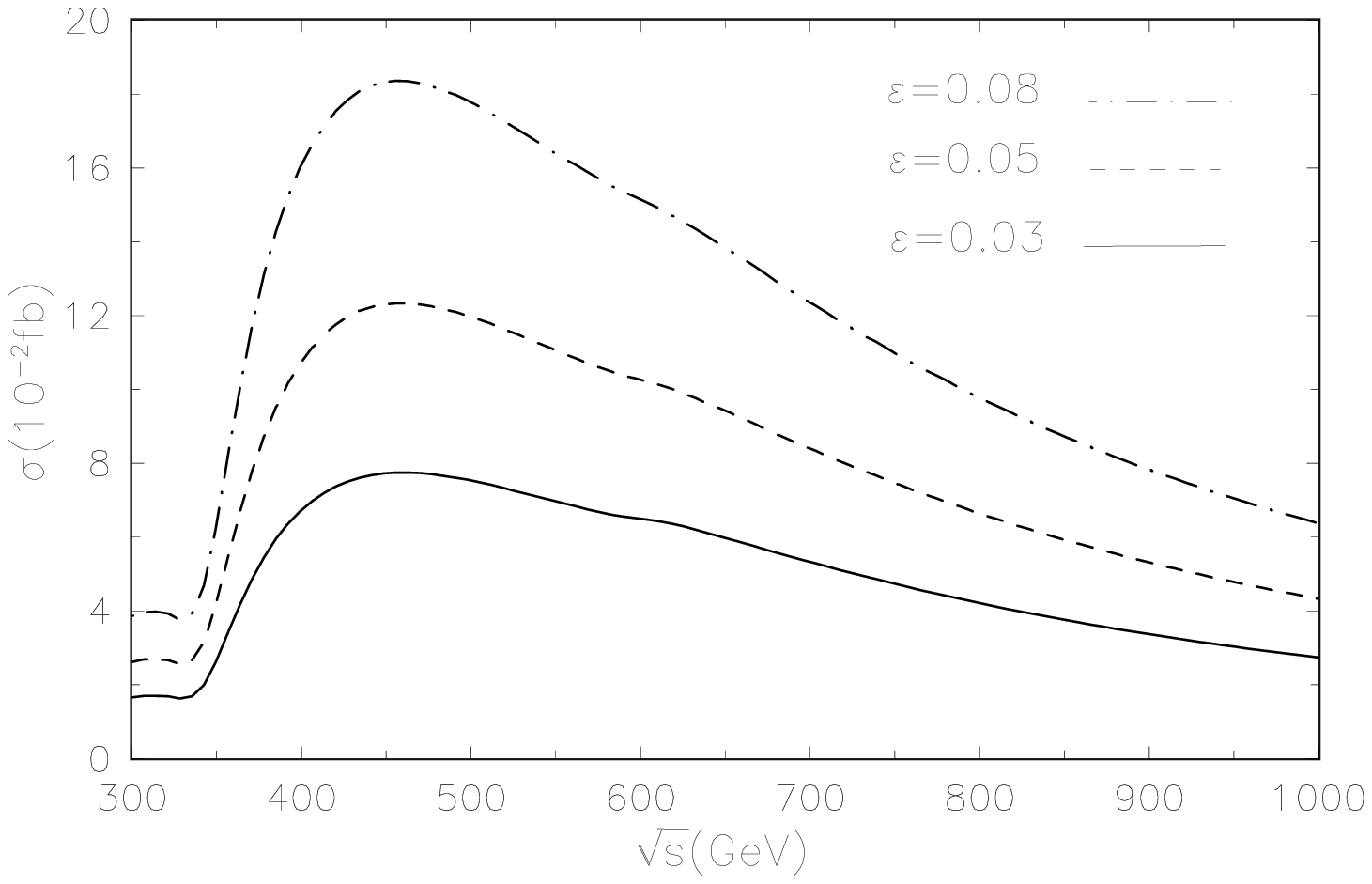}}\put(120,-10){Fig.4}
\end{picture}
\end{center}
\end{figure}
\newpage
\begin{figure}[pt]
\begin{center}
\begin{picture}(250,200)(0,0)
\put(-50,0){\epsfxsize120mm\epsfbox{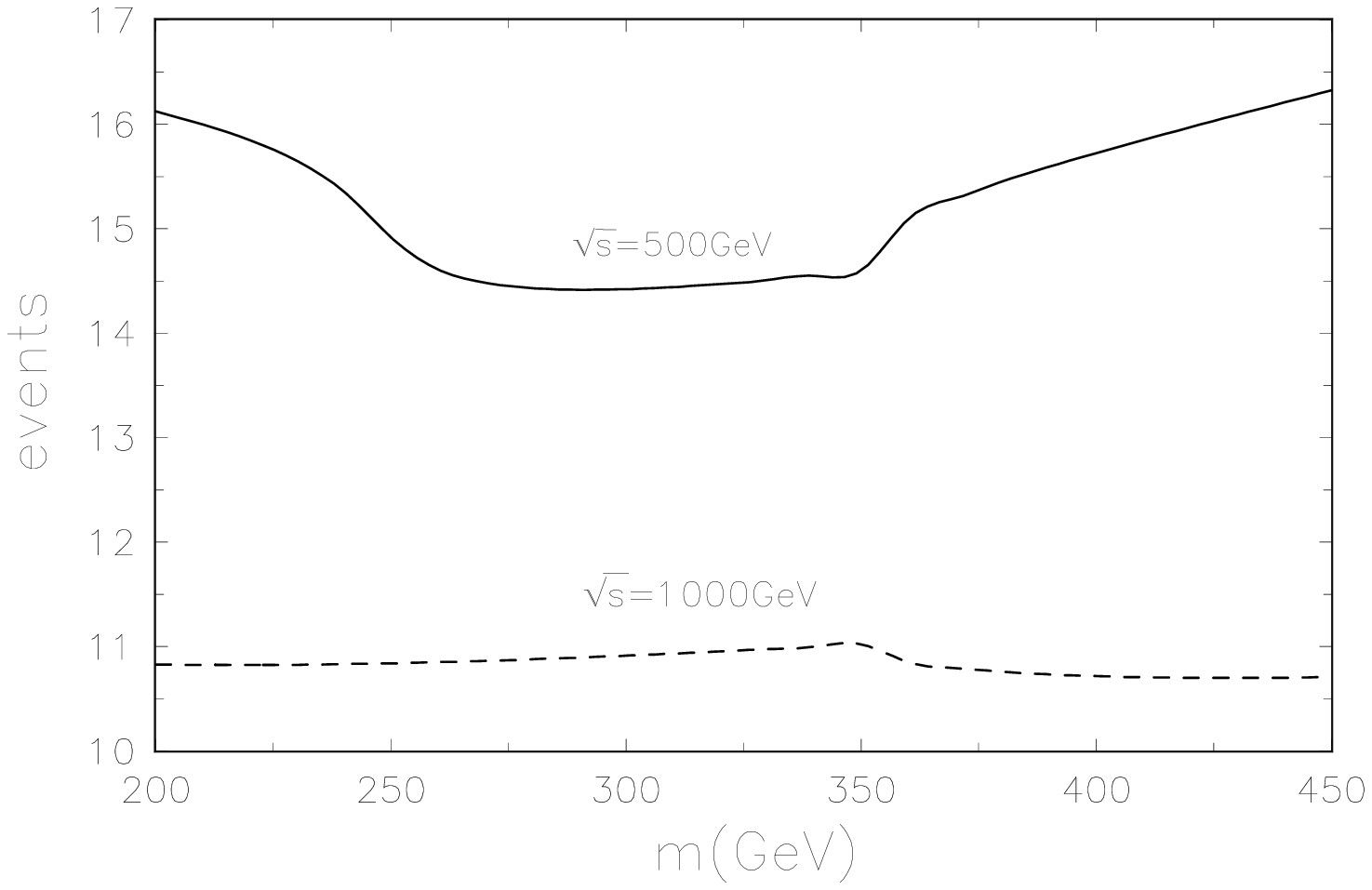}}\put(120,-10){Fig.5}
\end{picture}
\end{center}
\end{figure}
\begin{figure}[pt]
\begin{center}
\begin{picture}(300,250)(0,0)
\put(-50,-100){\epsfxsize160mm\epsfbox{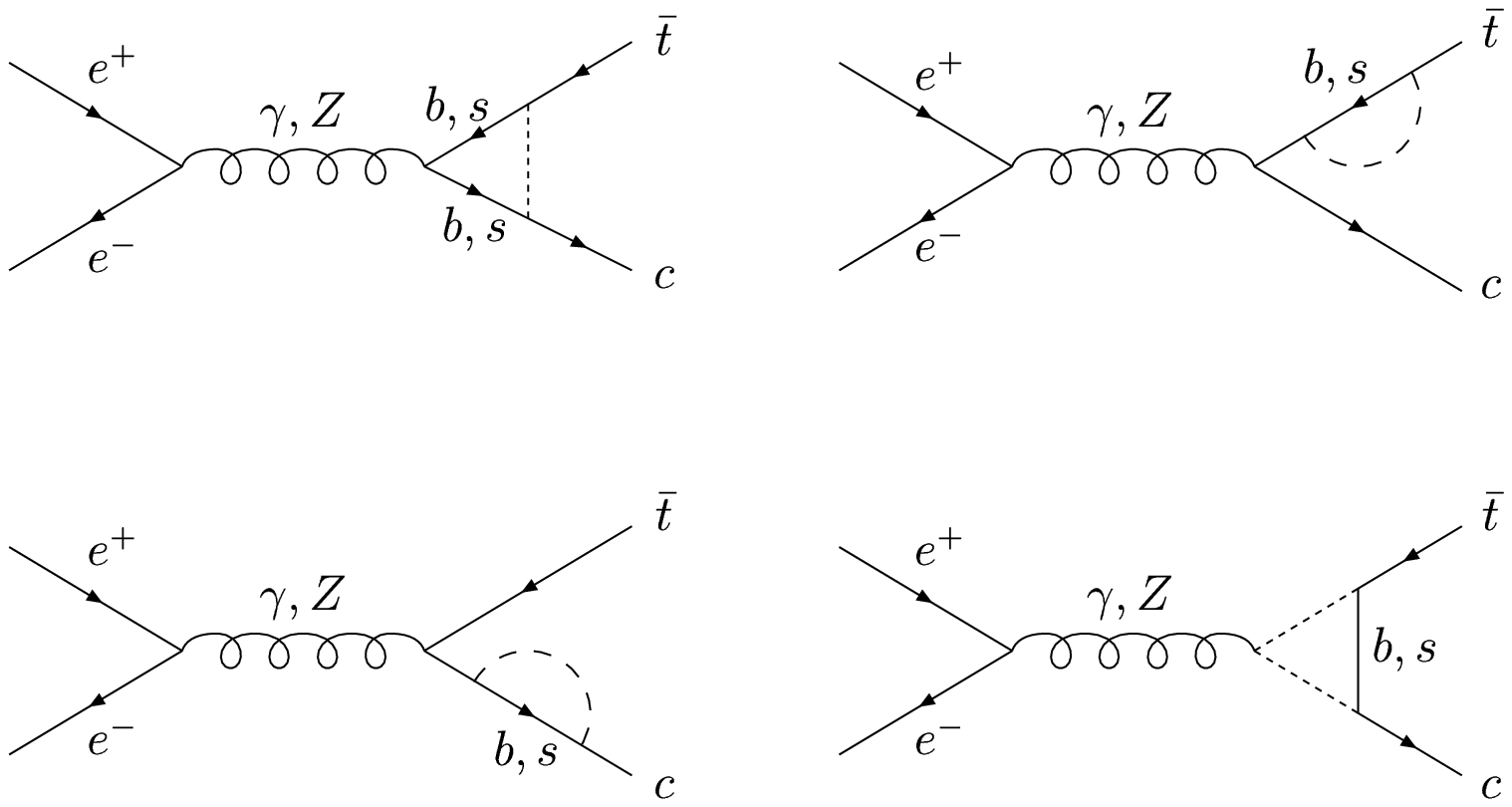}}\put(150,50){Fig.6}
\end{picture}
\end{center}
\end{figure}
\end{document}